\documentclass[amsmath,amssymb,superscriptaddress,aps,a4paper,prb,twocolumn]{revtex4} 

\usepackage[dvips]{color}
\usepackage{graphicx}
\usepackage{dcolumn}
\usepackage{bm}

\begin{document}


\title{\textcolor{blue}{Singlet-triplet transition in a few-electron lateral In$_{.75}$Ga$_{.25}$As/In$_{.75}$Al$_{.25}$As quantum dot}} 



\author{F.Deon}
\email{f.deon@sns.it}
\author{V.Pellegrini}
\author{F.Carillo}
\author{F.Giazotto}
\author{L.Sorba}
\affiliation{NEST, Istituto Nanoscienze-CNR and Scuola Normale Superiore, I-56127 Pisa, Italy}
\author{G.Biasiol} 
\affiliation{CNR-IOM, Laboratorio TASC, Area Science Park, I-34149 Trieste, Italy}
\author{F.Beltram}
\affiliation{NEST, Istituto Nanoscienze-CNR and Scuola Normale Superiore, I-56127 Pisa, Italy}\date{\today}   

\begin{abstract}
The magnetic-field evolution of Coulomb blockade peaks in lateral In$_{.75}$Ga$_{.25}$As/ In$_{.75}$Al$_{.25}$As quantum dots in the few-electron regime is reported. Quantum dots are defined by gates evaporated onto a 60 nm-thick hydrogen silsesquioxane insulating film. A gyromagnetic factor $g^*\approx 4.4$ is measured via zero-bias spin spectroscopy and a transition from singlet to triplet spin configuration is found at an in-plane magnetic field $B = 0.7$ T. This observation opens the way to the manipulation of singlet and triplet states at moderate fields and its relevance for quantum information applications will be discussed.

\end{abstract}


\maketitle 


Renewed interest in quantum dots (QDs) defined by lateral electrostatic gates stems from the relevance of these systems in the field of quantum information processing. Indeed QDs containing few electron spins can be operated as qubits\cite{PhysRevA.57.120,RevModPhys.79.1217}. State initialization, measurement and quantum-gate operation were experimentally demonstrated\cite{petta_coherent_2005,Pioro-Ladriere2008}. Manipulation of the electron spin was obtained through magnetic fields or electrically by exploiting spin-orbit interaction\cite{nowack_coherent_2007}. Much of the experimental work performed so far is based on GaAs-based heterostructures, for which a well-established technology is available. On the other hand, InGaAs QDs with high In concentration are attractive systems for spin manipulation thanks to their high effective g-factor\cite{PhysRevB.49.14786} and strong spin-orbit coupling\cite{PhysRevLett.78.1335,PhysRevB.69.245324}. Their exploitation was hindered, however, by the absence of a sufficiently large Schottky barrier. Indeed, the latter is virtually absent in In$_{.75}$Ga$_{.25}$As-metal contacts, a fact that stimulated the use of these alloys for the investigation of proximity effects\cite{Taddei:RevMesHybr} in hybrid semiconductor/superconductor devices\cite{PhysRevB.78.052506,schapers:3575}. 
Moreover, few-electron QDs were realized in InAs-based nanowires\cite{csonka_giant_2008,PhysRevB.72.201307}.
\begin{figure}[htb]
	\centering
		\includegraphics[width=0.45\textwidth]{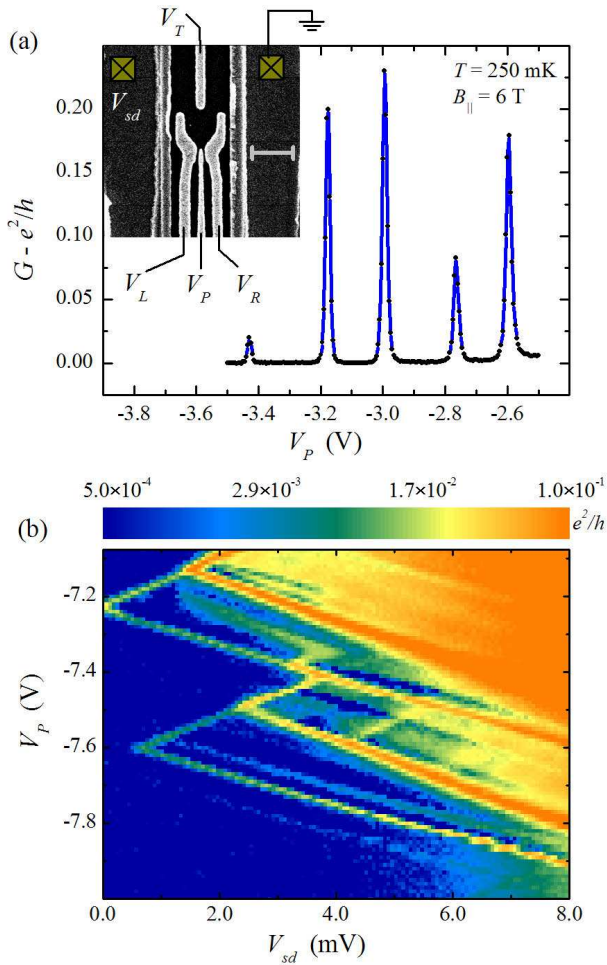}
	\caption{(Color online) (a) Zero bias differential conductance $G$ vs $V_P$ at 250 mK and 6 T (parallel field), measured with an AC excitation of $10$ $\mu$V. Inset: scanning electron micrograph of the device (the scale bar is 500 nm). (b) Stability plot, i.e., conductance versus $(V_P,V_{sd})$ at zero applied magnetic field and $250$ mK, plotted on a logarithmic scale. Very large addition energies $E_{add}\approx 6$ meV are observed, due to the small electron effective mass in In$_{.75}$Ga$_{.25}$As. }
	\label{Figure1}
\end{figure} 
Recently, the realization of In$_{.75}$Ga$_{.25}$As lateral QDs was reported by Sun \emph{et al.}\cite{sun:042114} using atomic layer deposition-grown hafnium oxide and by Larsson \emph{et al.}, with an approach that combines wet chemical etching and metal gating\cite{larsson:086101,larsson:192112}, following insulation by a 500 nm-thick dielectric layer.

In this letter we report magnetotransport measurements on lateral In$_{.75}$Ga$_{.25}$As QDs defined on an In$_{.75}$Ga$_{.25}$As/In$_{.75}$Al$_{.25}$As heterostructure, with gate insulation obtained with a 60 nm-thick layer of hydrogen silsesquioxane (HSQ)\cite{schapers:122107}. We report a crossover from a singlet to a triplet spin state that occurs with an in-plane  magnetic field of $0.7$ T. The observation that singlet and triplet spin states can be realized and manipulated in these lateral QDs at moderate fields offers promising venues for their exploitation in studies of spin physics and for quantum information processing.

\begin{figure}[t]
	\includegraphics[width=0.45\textwidth]{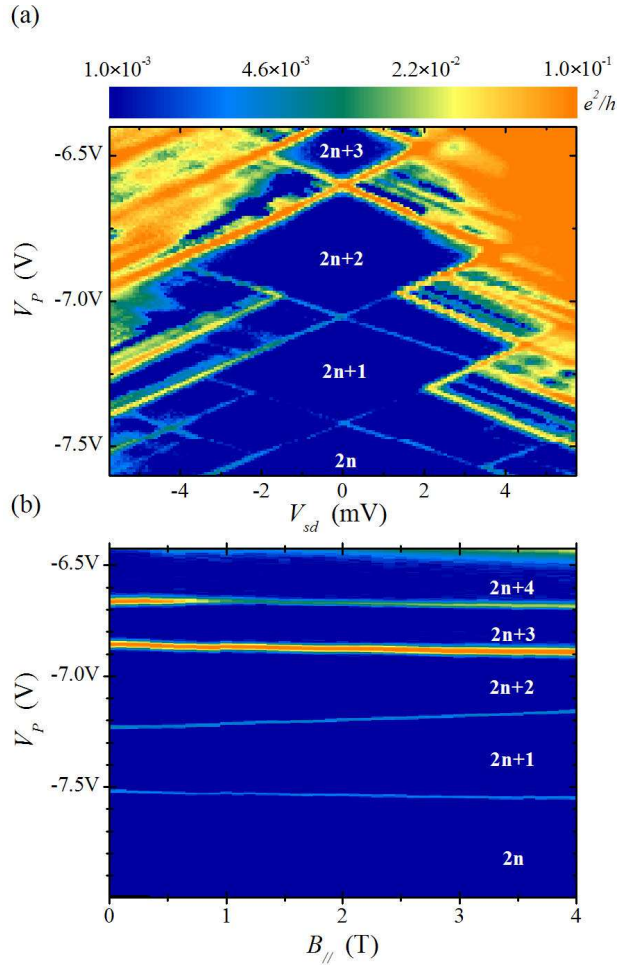}
	\caption{(Color online) (a) Stability plot at $B_{//}=0$, and (b) evolution of the zero-bias conductance as a function of the in-plane field for the same gate voltages. Charge rearrangement between the two measurements caused a small shift in $V_P$ between the two plots. Both datasets were taken at $250$ mK.}
	\label{Figure2}
\end{figure}

We employ two dimensional electron gases (2DEGs) confined in metamorphic In$_{.75}$Ga$_{.25}$As/ In$_{.75}$Al$_{.25}$As heterostructures grown on undoped (001) GaAs substrates by solid-source molecular beam epitaxy\cite{capotondi:702,capotondi/TSF2005}. A $\approx$1 $\mu$m-thick In$_x$Al$_{1-x}$As ``virtual crystal'' with stepwise increasing indium concentration ($x=0.15$ to $0.75$) is grown between the GaAs substrate and the active region in order to match the GaAs lattice constant to that of In$_{.75}$Ga$_{.25}$As and In$_{.75}$Al$_{.25}$As. Our heterostructure is designed as follows: a Si $\delta$-doped layer is followed by a 11 nm In$_{.75}$Al$_{.25}$As spacer, then by a 15 nm In$_{.75}$Ga$_{.25}$As quantum well and finally by a 45 nm In$_{.75}$Al$_{.25}$As barrier. The structure is capped with 5 nm of In$_{.75}$Ga$_{.25}$As. Shubnikov-de Haas measurements at 250 mK on the 2DEG yield a single occupied subband with density $n=6.25\times 10^{11}$ cm$^{-2}$ and mobility $\mu=2.0 \times 10^5$ cm$^2/$Vs.

A scanning electron micrograph of the sample is shown in the inset of Fig. \ref{Figure1}(a). Sample fabrication starts with patterning a 600 nm wide HSQ strip on the substrate by electron beam lithography (EBL)\cite{Lauvernier2004177,maex:8793}: film thickness is 60 nm. The central mesa region, aligned with the dielectric strip, is then defined by EBL and wet etching in a H$_3$PO$_4-$H$_2$O$_2$ aqueous solution using a PMMA mask. Etching depth is 90 nm, ensuring removal also of the underlying $\delta$-doping layer. Top metal gates are later patterned by EBL and liftoff on the dielectric. Finally ohmic side-contacts to the 2DEG are deposited by DC magnetron sputtering and liftoff. Low contact resistance was achieved by \emph{in situ} argon plasma cleaning thanks to the negligible Schottky barrier. The measured resistance of the 2.5 $\mu$m-wide mesa strip is 160 $\Omega$.

\begin{figure}[t]		
\includegraphics[width=.45\textwidth]{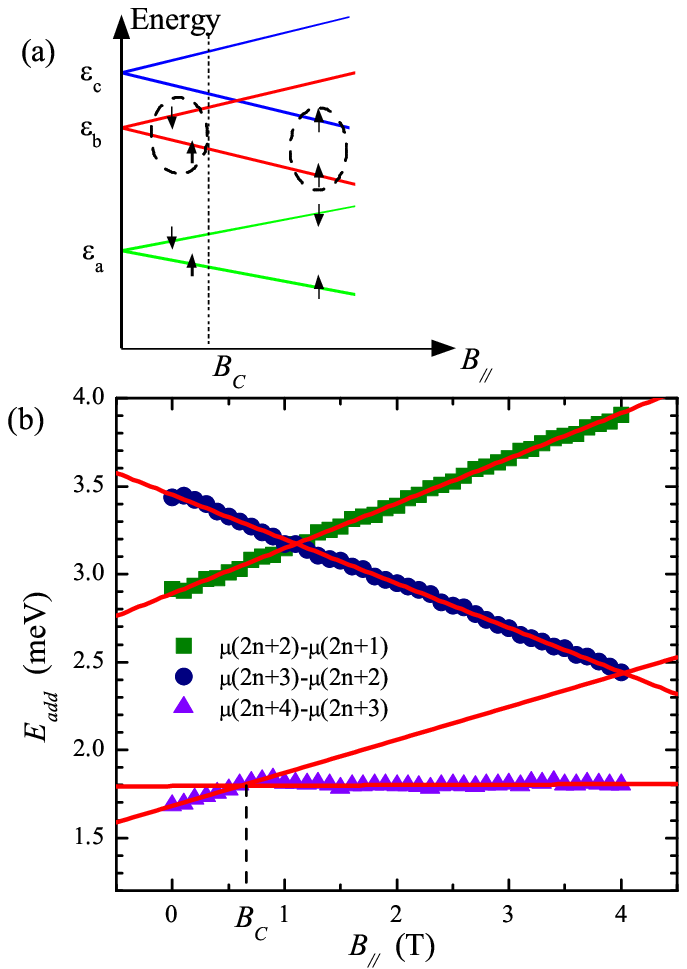}
	\caption{(Color online) (a) Sketch of the energy levels crossing due to Zeeman splitting. At $B_{//}=B_C$ the Zeeman and exchange energies cause the formation of a triplet spin state. (b) Addition energies extracted from Fig. \ref{Figure2}(b). The peak positions were determined by fitting the $G(V_P)$ for each value of $B_{//}$ with the lineshape for a single-level QD resonance\cite{PhysRevB.44.1646}.}
	\label{Figure3}
\end{figure} 
 
Electrical measurements are performed in a filtered $^3$He cryostat with 250 mK base temperature by lock-in technique. A DC $V_{sd}$ voltage plus a $10 \mu$V AC excitation are applied to the source electrode, with the drain electrode connected to the system ground through the input resistance of a low-noise current amplifier. Figures \ref{Figure1}(a) and (b) show representative low-temperature transport characteristics, typical of a QD in the Coulomb blockade regime\cite{0034-4885-64-6-201}: as $V_{sd}$ and $V_P$ are swept, diamond-shaped regions of reduced conductance are found [Figs. \ref{Figure1}(b) and \ref{Figure2}(a)]. Within each diamond conductance is strongly suppressed by Coulomb repulsion, and the island is populated by a constant number of electrons. At the edges of these diamonds the electrochemical potential in one of the leads is aligned to a QD resonance, thus lifting the blockade. The half width of the n$^{th}$ diamond is a direct measurement of the addition energy $E_{add}(n)$. Although we are not able to completely deplete the dot, the significant fluctuations in $E_{add}$, together with the large energy spacings and $E_{add}$ are typical of a dot in the few-electron regime\cite{0034-4885-64-6-201}.

The ratio of $E_{add}(n)$ to zero-bias peak separations allows us to estimate an average lever arm $\alpha\approx 9\times 10^{-3}$, that varies\cite{Note1} by $\approx 10\%$ in the range of plunger voltages shown in Fig. \ref{Figure2}. Additional lines that run parallel to diamond edges appear when $V_{sd}$ is sufficiently large to allow conduction through excited QD states. Data in Figs. \ref{Figure1}(b) and \ref{Figure2}(a) allow us to estimate single-particle energy spacing exceeding 1 meV.

The spin filling sequence of a QD can be inferred from the chemical-potential evolution $\mu(n,B)$ under parallel magnetic field $B_{//}$. Figure \ref{Figure2}(b) shows the measured zero-bias peaks in a $V_P - B_{//}$ plane, while the extracted $E_{add}(n)$ are reported in Fig. \ref{Figure3}(b). Data are consistent with the energy diagram depicted in Fig. \ref{Figure3}(a): single-particle orbital states $a, b, c$, with energies $\varepsilon_a,\varepsilon_b$ and $\varepsilon_c$ are non degenerate and all empty when $2n$ electrons are in the dot. At magnetic fields $B < B_C \approx 0.7$ T we observe the usual \emph{antiferromagnetic} filling, i.e. electrons are added with alternating spin orientation: $(a,\uparrow)$, $(a,\downarrow)$, $(b,\uparrow)$, $(b,\downarrow)$. Addition energies, in this case, are given by
\begin{eqnarray}
&&\mu_{2n+2}-\mu_{2n+1}=U+|g^*_a|\mu_B B\label{addition1} \\
&&\mu_{2n+3}-\mu_{2n+2}=U+(\varepsilon_b-\varepsilon_a)-\frac{|g^*_a|+|g^*_b|}{2}\mu_B B\label{addition2} \\
&&\mu_{2n+4}-\mu_{2n+3}=U+|g^*_b|\mu_B B,\label{addition3}
\end{eqnarray}
where $U$ is the charging energy, $g^*_{a,b}$ are the level-dependent effective $g$-factors, and $\mu_B$ is the Bohr magneton. The last two added electrons occupy the $\varepsilon_b$ level with opposite spins and form a singlet. 

As the magnetic field increases, the Zeeman energy brings levels $(b,\downarrow)$ and $(c,\uparrow)$ closer. As previously seen in similar QDs\cite{PhysRevLett.77.3613,larsson:192112}, the exchange energy $K_{bc}$ can be of the order of a fraction of meV and favors parallel spin filling. When $B>B_C$, $|K_{bc}|>(\varepsilon_c-\varepsilon_b)-(|g^*_c|+|g^*_b|)\mu_B B /2$ and transition to a triplet state takes place. In this configuration Eq. \ref{addition3} must be replaced by
\begin{equation}
\mu_{2n+4}-\mu_{2n+3}=U+(\varepsilon_c-\varepsilon_b)-|K_{bc}|+\frac{|g^*_b|-|g^*_c|}{2}\mu_B B,
\label{cambia}
\end{equation}
where $g^*_c$ is the effective $g$-factor of level $c$. For occupation numbers $2n+1$ and $2n+3$ the addition energy equals the charging energy $U$. The observed marked dependence of the latter on $V_P$ is a consequence of our dot design and of the high 2DEG density. By fitting (\ref{addition1}) and (\ref{addition2}) to the data in Fig. \ref{Figure3}(b), we obtain the effective $g$ factor moduli $|g^*_a|=4.4$ and $|g^*_b|=4.3$. For $B_{//}>0.7$ T the addition energy $\mu(2n+4)-\mu(2n+3)$ is constant, therefore $|g^*_c|=|g^*_b|=4.3$.

In conclusion, we have shown that it possible to reproducibly obtain metal-gate insulation and stable QDs on In$_{.75}$Ga$_{.25}$As/In$_{.75}$Al$_{.25}$As heterostructures using EBL-patterned HSQ. Thanks to the large value of the gyromagnetic factor $g^*\approx 4$ coincidence between singlet and triplet states was observed at low magnetic field values. This technique can open the way to the application of In$_{.75}$Ga$_{.25}$As lateral QDs for spin manipulation and quantum computing architectures.

We acknowledge S. De Franceschi for helpful discussions and the E.U. project HYSWITCH (grant No. FP6-517567) for financial support. F.G. acknowledges the NanoSciERA ``NanoFridge" project for partial financial support.


\end{document}